# From archive to resource: bringing legacy data back to life at scale with large language models

**Vieira-Vieira C. H. [1#], Schneider Chernov I.[2#], Serra A.[2], Correa Marrero M.[2], Bhavani Mohan P. L.[2], Duchateau-Nguyen G. [3], Doktorova T. Y. [2*], Kreuchwig, A.[1*]**

[1]Bayer Pharma AG, Research & Development, Preclinical Development, Berlin, Germany.

[2]F. Hoffmann-La Roche Ltd, Predictive Modelling, pRED, Basel, Switzerland

[3]F. Hoffmann-La Roche Ltd, Computational Sciences Center of Excellence, gRED, Basel, Switzerland

# shared first author

*** Correspondence:**
Corresponding Authors
annika.kreuchwig@bayer.com
tatyana.doktorova@roche.com

**Abstract**

Pharmaceutical companies maintain vast archives of legacy preclinical study reports that contain study-context details, such as fasting, anesthesia, and laboratory instrumentation, that are rarely captured in structured data formats, yet they can correlate with clinical pathology parameters values and limit the reuse of historical data. This work demonstrates practical, scalable solutions to recover such descriptors from heterogeneous report texts using large language model (LLM)-based extraction pipelines developed independently by two industry teams. The pipelines identify and normalize metadata fields and return verbatim evidence with page and section reference, hasting human verification and iterative refinement. Evaluation of over 400 documents against expert-curated reference values showed high agreement across a broad set of metadata fields. Performance was strongest for consistently reported items and lower for descriptors subject to linguistic or report structure variability. Extraction time has shrunk from 25 minutes in the manual benchmark to 1.6 minutes. As a concrete biological demonstration, we show that fasting status extracted from report text was associated with species- and analyte-specific shifts in clinical chemistry parameters, illustrating how recovered metadata can improve selection of historical controls. Key strengths of the approach are scalability, traceability, and the ability to recover information missing from current industry standards. Main limitations are dependence on source-report quality and initial expert effort to define and refine schemas. These results establish how LLMs can convert heterogeneous, unstructured legacy reports into evidence-grounded, reusable and structured metadata, leading to better experimental comparability, and thus enabling reliable data pooling or informed virtual control selection.

## Introduction

Pharmaceutical companies hold large collections of legacy preclinical study data from toxicology, safety pharmacology, and pharmacokinetic programs accumulated over decades. When made findable, accessible, interoperable and reusable (FAIR) (Wilkinson et al., 2016), these data can support cross-study safety assessments, evidence-based portfolio decisions (Alharbi et al., 2023), and the reuse of historical information in line with the 3R principles of Replacement, Reduction, and Refinement (Pognan et al., 2021). One application in which reuse of historical data may directly contribute to the 3Rs is the construction of virtual control groups (VCGs), as proposed by the Innovative Health Initiative-funded "Developing and implementing VIrtual Control groups To reducE animal use in toxicology Research" (VICT3R) consortium (Steger-Hartmann et al., 2025).

VCGs use appropriately matched historical control data (HCD) to replace concurrent control animals in new studies (Steger-Hartmann et al., 2020). By providing an alternative to including concurrent control animals, VCGs have the potential to reduce animal use while maintaining an appropriate reference for the interpretation of treatment-related findings, thereby supporting the 3R principle of Reduction. Their utility, however, depends on the availability of HCD that are sufficiently comparable to the animals in the new study (Steger-Hartmann et al., 2020; Golden et al., 2024). Furthermore, factors such as vehicle composition, diet, fasting conditions, anesthetic protocols, animal supplier, test facility, and housing conditions may affect clinical pathology parameters variability and act as confounders in cross-study comparisons (Palazzi et al., 2024). In addition, other, as-yet-unidentified confounders may also contribute to variability (Gurjanov et al., 2023). Many such factors are not consistently represented in current structured preclinical data standards, like CDISC's Standard for Exchange of Nonclinical Data (SEND) (CDISC, 2021), and are instead locked away across different sections of heterogeneous study reports, making their systematic identification challenging. In effect, much of this information is dormant as it has been recorded, but it cannot be used.

Manually recovering this information requires subject-matter expertise and does not scale to the thousands of reports held by companies. Large language models (LLMs) can automate this process, and earlier work at Bayer and Roche showed that they can be used reliably at scale (Vieira-Vieira et al., 2025; Doktorova et al., 2026). To our knowledge, three questions remained open: whether this approach can reliably recover the contextual metadata that current preclinical data standards leave out, whether it holds up across different reports with different formats and conventions, and whether the recovered information is accurate enough to impact how historical data are analyzed.

Here, we address all three. Bayer and Roche independently developed LLM-based pipelines that process legacy study reports to extract contextual metadata relevant to VCG construction and not captured in SEND (CDISC, 2021), such as fasting and anesthesia before blood sampling. Each extracted value is linked to the paragraph it came from. The two pipelines find relevant text differently: Bayer's searches each report for the most relevant paragraphs, while Roche's applies natural language processing (NLP) to identify the report body before passing the text to the language model. Evaluated against expert-curated values across 244 Bayer and 202 Roche reports, both matched expert curation for most study metadata. Finally, we show that the recovered information has analytical value: fasting status, available only in report text, was associated with changes in clinical chemistry parameters in control animals, making it a candidate matching criterion for VCGs. Together, these results show that information lying dormant in legacy reports can be brought back to life, accurately and at scale, to improve the reuse of historical control data and support efforts to reduce animal use in toxicology.

## Methods

### System Architecture – PRINCE (Bayer)

Extraction of study metadata from preclinical toxicology study reports followed a schema-constrained information extraction approach, in which an LLM populated a predefined set of typed attributes directly from source documents. Each target attribute was treated as a named entity to be located, extracted, normalized, and grounded in the source text. The approach used to process unstructured text from study reports has been described previously (Vieira-Vieira et al., 2025). In brief, source study reports were pre-processed into text chunks and indexed in an OpenSearch vector store. For a given study, retrieval was restricted to the target study using a pre-filter, and relevant chunks were selected by similarity search within a retrieval-augmented generation (RAG) framework.

Attribute extraction was carried out with the GPT-5.4 model, served through Bayer's internal GenAI platform myGenAssist (Benaïche et al., 2025). A fixed system prompt instructed the model to act as an expert extraction algorithm, to extract only information explicitly supported by the provided text, and to return null for any attribute whose value could not be determined. The study report text chunks were supplied as the user context, and the model returned one structured object per schema group using structured-output (constrained decoding) mode defined by a Pydantic schema (Colvin et al., 2026). Because several attributes were often derived from the same chunks, attributes were organized into thematic schema groups. For fields left unresolved, a focused follow-up query was issued with a smaller retrieval window tailored to the individual attribute, and the structured model was re-invoked.

Finally, extracted values were reconciled with existing values in Bayer's internal database using one of three rules, depending on the field:

- Fields absent from the PRINCE database (AGE, AGEU, HOUSETYP, HUMIDTU, LIGHT, TEMPU, WATER, TRTV, EXVAMT, EXDOSFRM, RECSAC, SPLRCTRY, SPLRREG, TSTFNAM, EXENDTC, MANAESTH): the extracted value was accepted unconditionally.
- Fields where PRINCE curation was more reliable owing to controlled vocabularies or structured entry (SPECIES, EXDOSFRQ, EXDOSTXT, DIET, HOUSEGRP, BW, CL, FW, LB, PP, MA, MI, OM, VS, EG, CV, RE): the PRINCE value was retained when available, and the extracted value was used only as a fallback.

- Remaining fields present in both sources (STRAIN, GLPFL, HUMIDT, TEMP, ROUTE, DOSDUR, TRT, SPLRNAM, SPLRLOC, STDIR, EXSTDTC): the extracted value was used preferentially, and the PRINCE value substituted only when extraction returned no result.

**System architecture - Controlled Generation (Roche)**

The preprocessing and LLM-based extraction pipeline adapted a previously published study-report extraction architecture (Doktorova et al., 2026) for VCG metadata extraction (Figure 1). The workflow comprised report-core identification, parsing, and optional section selection, which is shown in Figure 1 for completeness, but was not used in this study. Report cores were identified using PyMuPDF (Jorj X. McKie, 2026) and PyPDF2 (Fenniak et al., 2024) and parsed using MinerU (Wang et al., 2024) or olmOCR (Poznanski et al., 2025). Detailed preprocessing, fallback logic, and parser selection are provided in supplementary material 3. Metadata was extracted using GPT-4.1 (OpenAI, 2025), with one call per parsed report section and metadata field. A Pydantic - based schema defined field-specific extraction instructions and output formats (Supplementary Figure 1). Fields were selected based on SEND, relevance to preclinical toxicity studies, and VICT3R historical-control requirements (EMA, 2026), with study-specific extensions relevant to enable cross-study comparison and VCG development. Complete field definitions, SEND mappings, extension-field rationale, and output-handling rules are provided in supplementary table 1 and supplementary material 3.

**Iterative human curation of reference values with LLM assistance**

In both the Bayer and Roche workflows, extraction pipelines were evaluated at the level of individual metadata fields against manually curated reference values. For every extracted content item, the pipeline returned the extracted value, the verbatim text on which the value was based, the page number of that text, and the name of the report section from which it originated. This traceability information accelerated human curation by reducing the need for manual searching. It also enabled curators to distinguish between two error categories:

- Incorrect source selection: the model based its answer on a report section that did not contain the requested information. These cases typically indicated preprocessing errors (exclusion of the title page and, with it, the test facility name and location) or insufficient detail in the initial task formulation. For example, by not specifying the required detail level about hematology assays and instruments.

- Incorrect interpretation of a correct source: the model identified the correct section but drew an incorrect conclusion from it. These errors were mainly attributable to subtle details in the prompt formulation for specific items and were addressed by refining the extraction instructions with domain experts.

This cycle of automatic extraction, human curation, and prompt refinement was repeated until all entities were manually validated for the reports labeled as the test dataset.

## Automatic evaluation with LLM-as-a-Judge

Correctness against the manually curated reference values was assessed automatically by an LLM-as-a-Judge. The judge model was provided with the extracted value, the curated value, and the description of the corresponding metadata field, and was instructed to classify each extracted value as "Correct", "Partially Correct", or "Incorrect". A value was marked "Correct" when it was verbatim or semantically identical to the curated value, "Partially Correct" when it partially matched the curated value (for example, "Oral" or "Intragastric" extracted when the curated value was "Oral Gavage"), and "Incorrect" when it differed from the curated value. The judge could additionally output "Not Extracted" when no value was found in the source document, "No Curation and Not Extracted" when no curated value existed and nothing was extracted, and "No Curation and Extracted" when no curated value existed but a value was nevertheless extracted.

An example of the instruction provided to the LLM-judge in the Bayer pipeline is given in supplementary material 1. From these categories, a common confusion matrix was defined and used to derive precision, recall, and F1 score:

- True Positives (TP): values that fully or partially matched the curated value.
- False Positives (FP): values that did not match the curated value, or values that were extracted although no curated value existed.
- True Negatives (TN): cases in which no curated value existed, and no value was extracted.
- False Negatives (FN): cases in which a curated value existed, but no value was extracted.

A curated value of "Not Available" indicated that the entity was absent from the study report and therefore could not have been legitimately extracted and any value returned for such a field was counted as a false positive. Representative examples for each category are shown in Table 1.

## Linear mixed-effects model

To illustrate how the extracted metadata can support biological analyses, we examined the association between fasting status and clinical chemistry values in monkeys and rats. Studies were excluded if fasting status was unclear or not found during extraction. Values were log-transformed because their distributions were right-skewed and their spread increased with the mean; fasting effects are therefore expressed as fold changes. For each of the 16 species-analyte combinations (2 species x 8 analytes), we fitted:

$$\log(Y_{ij}) = \beta_0 + \beta_{fast} \cdot F_j + \beta_{sex} \cdot S_{ij} + \beta_{spec} \cdot P_j + u_j + \varepsilon_{ij}\ ,\ u_j \sim \mathcal{N}(0, \sigma^2_u),\ \varepsilon_{ij} \sim \mathcal{N}(0, \sigma^2)$$

where $\beta_0$ is the intercept, $y_{ij}$ is the analyte value for animal *i* in study *j,* $F_j$ is a dummy variable for fasting status (1 = fasted, 0 = unfasted), $S_{ij}$ is a dummy for sex (1 = male, 0 = female), and $P_j$ is a dummy for specimen type (1 = plasma, 0 = serum). All fixed terms are main effects (coefficient × indicator variable) and no interaction terms were included in these models. $\beta_0$ is the intercept, for example the mean log analyte value in the reference group (unfasted, female, serum), and the coefficients $\beta_{fast}$, $\beta_{sex}$, and $\beta_{spec}$ represent the adjusted differences associated with fasting, male sex, and plasma specimens, respectively. $u_j$ is the random intercept for study j, representing study-level variation in the outcome, and $\varepsilon_{ij}$ is the animal-level residual. Models were fitted with restricted maximum likelihood using statsmodels v0.14.4 library, with Wald-based confidence intervals and p-values. Fasting p-values were corrected across the 16 tests using the Benjamini-Hochberg procedure (Benjamini, 1995). These 16 models were fitted within species and provide all reported fold changes, confidence intervals and q-values.

A separate model per analyte, fitted on the pooled species data with a fasting-by-species interaction, contributed only the interaction p-value. Given the limited number of unfasted studies per species (12 monkey, 9 rat), this test was interpreted as a qualitative indicator of species differences, rather than as evidence of equivalence when non-significant. Animal age at study start and test facility were assessed as study-level confounders and excluded: age was balanced across fasting arms, and facility was nearly collinear with fasting and therefore absorbed by the study random intercept. Each model thus estimates the fasting-associated shift in one analyte and species, adjusted for sex, specimen type and between-study variation.

## Results

### Bayer results

To assess the performance of the Bayer extraction pipeline, metadata for 205 studies were manually curated to serve as reference values. This set of studies was named the training set. The only selection criterion for the training set was the availability of a study report for studies donated by Bayer AG to the VICT3R consortium up to August 2025. The training set comprised 165 GLP and 40 non-GLP studies. The oldest study dated 1991, and the median exposure start date was 2015 (range: 1991–2021).

Manual extraction of all 44 metadata fields across 205 studies required approximately 86 hours, corresponding to 25 minutes per study, when performed by a subject matter expert familiar with the diverse report formats. Non-GLP studies were particularly challenging because some parameters were not reported and, therefore, could not be curated from the study report alone. Where possible, curation was supplemented by other company-internal metadata sources, such as study director records or LIMS systems. Consequently, in some cases, perfect agreement between manually curated reference values and automated extractions from study reports alone cannot be achieved, as the report was the only source for the latter.

To establish a baseline comparison, we first compared previously annotated values in Bayer AG's internal database for preclinical studies, PRINCE (Vieira-Vieira et al., 2025), against the manually curated reference values (Figure 2A and Supplementary Figure 2A). In the PRINCE database, recall (mean: 81.8 ± 28.2%; median: 98.2%), precision (mean: 92.1 ± 16.0%; median: 100.0%), and F1 score (mean: 82.5 ± 24.7%; median: 92.3%) varied substantially across metadata fields, reflecting historical variation in metadata collection and reporting standards. Information on animal housing conditions (HUMIDT, TEMP, DIET) and dosing frequency (EXDOSFRQ) was often not present. In addition, many exposure start dates (EXSTDTC) were incorrect, likely due to the mismatched definitions for experimental start and exposure start in legacy systems from which PRINCE originally sourced data.

Next, the extraction pipeline was applied to study reports from all 205 studies in the training set. Extraction was performed iteratively: output from each round was manually reviewed and used to refine field-specific prompts. Early prompts used generic instructions, whereas later prompts incorporated more precise guidance on likely report sections and expected wording. Single-shot examples derived from previous iterations were also added to harmonize outputs. The final prompt used for the STRAIN metadata field is provided in supplementary material 2.

Automated extraction of all 44 metadata fields with the final prompts required 5.5 hours, corresponding to 1.6 minutes per study, and was limited mainly by LLM inference latency and extracted metadata fields were compared against the manually curated reference values (Figure 2B and Supplementary Figure 2B). Across the 28 metadata fields also available in the PRINCE database, average recall (mean: 99.8 ± 0.5%; median: 100.0%), precision (mean: 97.9 ± 2.4%; median: 98.5%), and F1 score (mean: 98.8 ± 1.3%; median: 99.3%) were markedly higher than the PRINCE baseline. Recall, precision, and F1 score across all 44 extracted metadata fields were also high (means: 95.8 ± 11.8%, 96.2 ± 10.1%, and 95.6 ± 11.2%, respectively; medians: 99.8%, 98.5%, and 98.7%). Importantly, metadata fields absent from or poorly annotated in the PRINCE database, including HUMIDT, TEMP, DIET, EXDOSFRQ, and EXSTDTC, were successfully extracted from study reports.

Performance varied across metadata fields. Extraction of anesthesia method (MANAESTH), a non-standard SEND metadata field that was often not reported in study reports, was substantially weaker than for other fields, likely because reporting was inconsistent (recall: 55.7%, precision: 32.5%). Animal supplier country (SPLRCTRY) was frequently not extracted (recall: 35.5%) but was usually correct when extracted (precision: 93.4%). Animal age and age units (AGE and AGEU) also showed lower performance, consistent with the frequent absence of this information from study reports.

To assess generalizability, we further evaluated 39 studies that were not included in the training set (Figure 2C and 2D, and Supplementary Figure 2C and 2D). This set of studies was named the validation set and included 20 GLP and 19 non-GLP studies. The median exposure start date was 2021 (range: 2009–2023). As before, metadata fields were manually extracted to serve as reference values. The extraction pipeline was then applied using the final prompts developed on the training set and extracted metadata fields were merged with previously available PRINCE annotations.

On the validation set, extraction performance across the 28 PRINCE-comparable metadata fields remained comparable to the training set, with recall of 99.1 ± 2.2% (median: 100.0%), precision of 95.1 ± 7.1% (median: 97.4%), and an F1 score of 97.0 ± 4.6% (median: 98.7%). The modest reduction in precision relative to the training set was likely attributable to fields with no extractable report value or no available curated reference value, consistent with the higher proportion of non-GLP studies in the validation set (19/39 versus 40/205). These results support the generalizability of the extraction approach.

**Roche results**

The Roche VCG extraction use case of 202 studies covered 10 metadata fields. Absolute TP, FP, TN, and FN counts are visualized in Figure 3A. For Test Facility Name and Test Facility Location, the number of evaluated instances (n = 220) exceeded the number of reports (n = 202), because some studies involved multiple facilities, each scored independently. By contrast, the additional instance for Fasting Status Prior Blood Collection represented a false-positive extraction.

The Roche extraction pipeline showed strong overall performance across evaluated metadata fields, with several fields reaching ceiling-level scores (Figure 3B). Test Facility Name and Test Facility Location achieved 100% accuracy, recall, precision, and F1 score. These fields were typically reported on the report cover page or a dedicated test-site information section, which likely contributed to reliable extraction. These high-performing metadata fields tend to be reported in consistent or formulaic language.

Performance decreased for fields whose wording and surrounding context varied more substantially across CROs and report formats. Anesthesia-related metadata fields showed very high recall (100% for Anesthetic Prior Blood Collection and 99% for Anesthetic Prior Death) but lower precision (73% and 72%, respectively). This indicates that the model identified nearly all relevant information but also produced false positives. One recurrent error pattern was a failure to distinguish anesthetic agents from euthanasia agents when reports listed them without explicitly differentiating their purpose.

The weakest performance was observed for Animal Supplier Name (F1 score: 79%) and Hematology Instrument (F1 scores: 85%). For Animal Supplier Name, errors likely reflected variability in how suppliers were referenced and whether supplier information appeared in report sections retained during preprocessing. Additional ambiguity arose when reports described the geographic origin of the animals but did not provide a supplier identity. For example, in references to “Mauritian cynomolgus monkeys”, the LLM incorrectly interpreted the geographic origin (Mauritius) as the animal supplier. For Hematology Instrument, errors suggest that the extraction instructions did not always compensate for domain-specific ambiguity, such as distinguishing an instrument from associated software or specifying an existing model of the instrument that exists but was not used in this study.

Overall, the Roche extraction pipeline achieved high reliability for structured and consistently expressed metadata fields, whereas performance was lower for fields with greater linguistic variability or domain-specific ambiguity.

**Fasting status as a relevant selection criterion for virtual control groups**

To demonstrate the utility of metadata extracted from study reports for VCG construction, we examined the association between Fasting Status Prior Blood Collection and eight clinical chemistry analytes: alanine aminotransferase (ALT), alkaline phosphatase (ALP), aspartate aminotransferase (AST), calcium, cholesterol, glucose, phosphate, and triglycerides. Only cynomolgus monkeys and Wistar Han rats had enough fasted and unfasted studies for analysis: 44 fasted and 12 unfasted monkey studies, and 31 fasted and 9 unfasted rat studies (Figure 4A). Other species and/or strains were excluded because too few studies were available in one or both groups.

Fasting was associated with analyte- and species-specific shifts in clinical chemistry parameters (Figures 4B and C). As expected, after FDR correction, fasting was associated with lower glucose in cynomolgus monkeys (fold change: 0.87; 95% CI: 0.79–0.96; $q = 0.033$), lower triglycerides in Wistar Han rats (fold change: 0.59; 95% CI: 0.41–0.85; $q = 0.033$), and higher phosphate in cynomolgus monkeys (fold change: 1.12; 95% CI: 1.04–1.20; $q = 0.033$). The fasting effect differed by species for triglycerides (interaction $p = 0.009$) and phosphate (interaction $p = 0.037$), supporting the decision to model species separately.

ALT, AST, calcium, and cholesterol showed no significant fasting-associated shift in either species. Rat ALP decreased by 25%, but this effect did not remain significant after FDR correction. Fasting therefore shifted values for three of eight analytes, and the affected analytes differed between species, making fasting status a relevant matching criterion for glucose and phosphate in cynomolgus monkeys and for triglycerides in Wistar Han rats. For the remaining analytes, we found no evidence that matching on fasting status is required, although the small number of unfasted studies does not exclude smaller effects.

It must be noted that fasting was nearly collinear with test facility. Therefore, facility-level differences, such as the analytical platform used, cannot be separated from the fasting effects, and the results, although physiologically justifiable, should be interpreted as associations.

## Discussion

Pharmaceutical companies often revisit historical study data to inform contemporary drug development (Salami et al., 2025; Seedat et al., 2025; Warner et al., 2026). In this study, Bayer and Roche independently developed LLM-based pipelines to recover contextual metadata from heterogeneous historical preclinical study reports. Both approaches achieved high agreement with expert curation (Figure 2 and 3), showing that metadata previously accessible only through costly, time-consuming manual review can be extracted at scale.

Preclinical data is increasingly pooled across companies to develop predictive models and improve translation of findings between humans, nonclinical species, and in-vitro systems (Sanz et al., 2017; Pognan et al., 2021; FDA, 2025; Steger-Hartmann et al., 2025). Missing study descriptors are particularly relevant for VCGs construction (Gurjanov et al., 2023, Palazzi et al., 2024), as these descriptors are used to select suitable historical control animals (Steger-Hartmann et al., 2025). The value of metadata extraction lies not only in its accuracy, but also in whether the recovered metadata is scientifically informative. We showed here that fasting status before blood collection was associated with significant differences in clinical chemistry parameters values (Figure 4). Our findings therefore support capturing fasting status in structured form in future versions of preclinical data standards, consistently with previous observations that current standards do not capture all relevant study-level metadata (Briggs, 2017; Pognan et al., 2021; van Mulligen et al., 2025).

### Methodological lessons, limitations, and outlook

Both extraction architectures from Bayer (Vieira-Vieira et al., 2025) and Roche (Doktorova et al., 2026) achieved similarly strong performance. Most false positives and missed values arose from linguistic variability, domain-specific ambiguity, and differences in reporting practices across test facilities and time periods, highlighting the influence of reporting heterogeneity on extraction quality. Altogether, these support the notion that ensuring robust parsing and document access and building pragmatic validation workflows, yield greater benefit than optimizing the extraction architecture (Li et al., 2024, 2025).

In both settings, extraction issues could be mitigated through iterative schema refinement, clearer extraction instructions, and targeted prompt updates. The fact that both pipelines retain verbatim evidence and report location is particularly important in this context, because it enables efficient human review and creates a foundation for automated verification and retry-based correction when outputs are uncertain (Dao et al., 2025; Wu et al., 2025).

One bottleneck in our approaches is adaptation to new domains. While the approaches are extensible, defining metadata schemas and extraction instructions still requires manual expert input (Hein et al., 2025). Emerging methods for automated performance assessment, prompt optimization, and agent-based support for requirement elicitation and schema drafting may help reduce this burden (Toro et al., 2024; Zhao et al., 2024; Khanmohammadi et al., 2025; Zhang et al., 2025).

In conclusion, information that has lain dormant in preclinical study reports can be unlocked and leveraged for new research purposes: LLM-based extraction recovers it accurately, at scale and with traceable evidence. Such metadata strengthens the evidence base for translational and regulatory decisions and supports efforts to reduce animal use.

**Conflict of Interest**

C.H.V.-V., A.K. are employed by Bayer AG. I.S.C., A.S., M.C.M, P.L.B.M., G.D.N, T.Y.D. are employed by Roche. The authors declare no further conflicts of interest.

**Author Contributions**

CHVV: Writing – original draft, review & editing, Methodology, Investigation, Conceptualization, Data Curation.

ISC: Writing – original draft, review & editing, Methodology, Investigation, Conceptualization; Data Curation.

AS: Writing – original draft, review & editing, Methodology, Formal Analysis, Data Curation.

MCM: Writing - original draft, review & editing, Investigation, Formal Analysis, Supervision

PLBM: Writing - original draft, review & editing, Data Curation

GDN: Writing - original draft, review & editing, Methodology, Project administration, Conceptualization

TYD: Writing – original draft, review & editing, Investigation, Conceptualization, Data Curation

AK: Writing – original draft, review & editing, Investigation, Project administration, Conceptualization.

**Funding**

VICT3R is supported by the Innovative Health Initiative Joint Undertaking (IHI JU) under grant agreement No 101172693. The JU receives support from the European Union's Horizon Europe research and innovation programme and COCIR, EFPIA, Europa Bío, MedTech Europe, and Vaccines Europe and Instem Scientific Limited. This project is funded by the European Union, the private members, and those contributing partners of the IHI JU. Views and opinions expressed are, however, those of the author(s) only and do not necessarily reflect those of the aforementioned parties. Neither of the aforementioned parties can be held responsible for them.

The author(s) declare that no further financial support was received for the research, authorship, and/or publication of this article.

**Acknowledgments**

The authors gratefully acknowledge the invaluable contributions of the PRINCE Project Team from Bayer and Dr. Dragomir Ivanov Draganov from Roche.

The authors declare that generative AI was used in the creation of this manuscript. *MyGenAssist* (internal Bayer Tool) was utilized solely for the purpose of enhancing language and readability. Internal ChatGPT (GPT-5.6 Sol) at Roche was used to prepare Figures 1 and Supplementary Figure 1.

## Data availability statement

The original contributions presented in the study are included in the article/supplementary material, further inquiries can be directed to the corresponding author.

## Figure Legends

Figure 1 – Automated extraction and standardization pipeline for study report metadata. The workflow converts PDF reports into structured datasets through: (1) preprocessing, including report core extraction and parsing to Markdown; (2) optional section selection; (3) LLM-based extraction using a predefined schema; and (4) expert curation and quality review. Feedback from expert review was used to refine preprocessing, section extraction, and schema definitions.

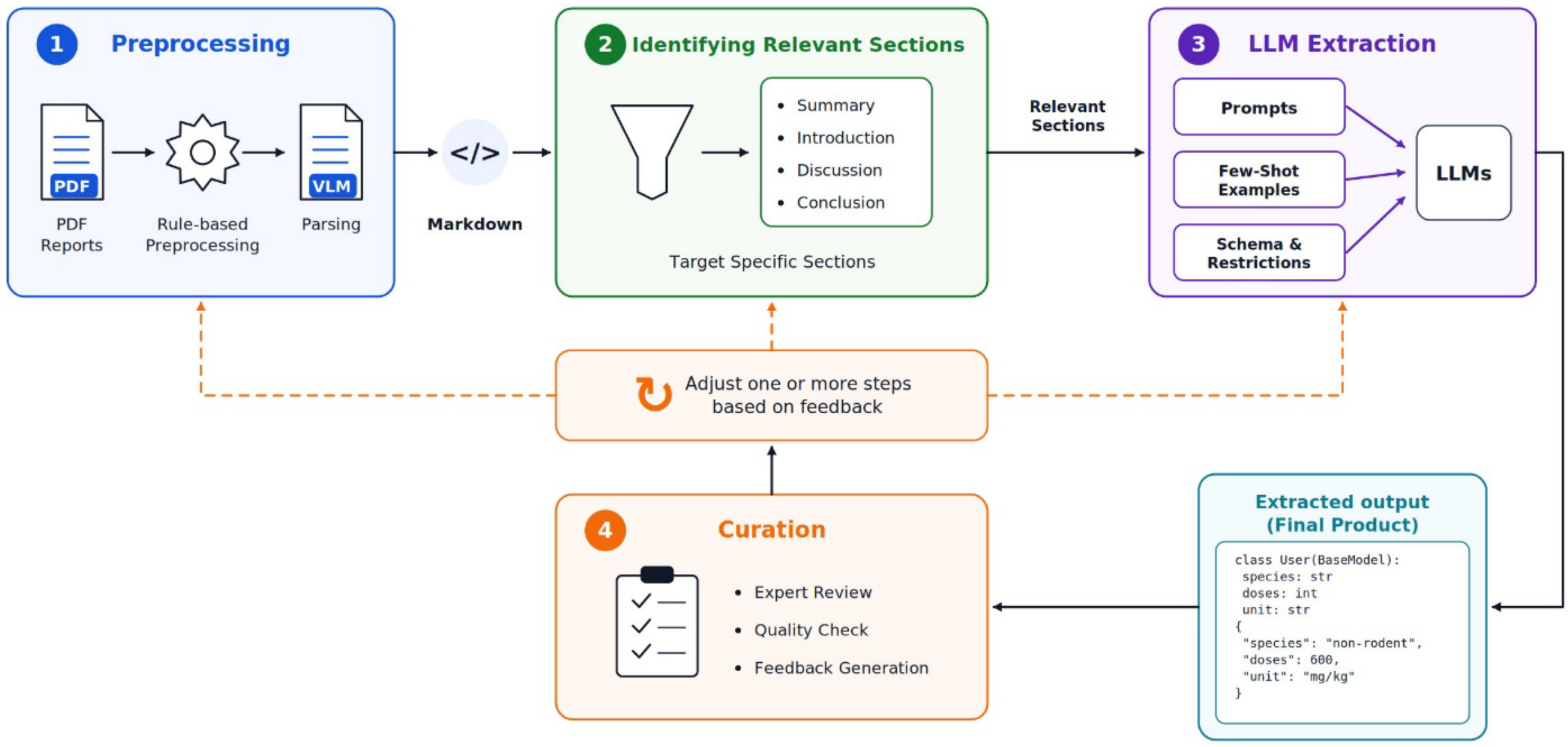

Figure 2 – Per-field confusion matrix benchmarked against manually curated metadata fields. Training set (n = 205), PRINCE baseline (A); Training set (n = 205), PRINCE plus extraction pipeline combined output after rule-based merging (B); Validation set (n = 39), PRINCE baseline (C); and validation set (n = 39), PRINCE plus extraction pipeline combined output after rule-based merging (D). In combined panels (B and D), bold field labels indicate metadata fields not present in PRINCE.

## Training set (n = 205 studies) Validation set (n = 39 studies)

**A** PRINCE baseline

| | TP | FP | TN | FN | Precision | Recall | F1 |
|---|---|---|---|---|---|---|---|
| SPECIES | 205 | 0 | 0 | 0 | 1.000 | 1.000 | 1.000 |
| STRAIN | 205 | 0 | 0 | 0 | 1.000 | 1.000 | 1.000 |
| GLPFL | 163 | 30 | 0 | 12 | 0.845 | 0.931 | 0.886 |
| HOUSEGRP | 157 | 48 | 0 | 0 | 0.766 | 1.000 | 0.867 |
| HUMIDT | 38 | 12 | 11 | 144 | 0.760 | 0.209 | 0.328 |
| TEMP | 39 | 11 | 10 | 145 | 0.780 | 0.212 | 0.333 |
| DIET | 38 | 3 | 7 | 157 | 0.927 | 0.195 | 0.322 |
| ROUTE | 204 | 0 | 0 | 1 | 1.000 | 0.995 | 0.998 |
| EXDOSFRQ | 35 | 2 | 0 | 168 | 0.946 | 0.172 | 0.292 |
| EXDOSTXT | 192 | 13 | 0 | 0 | 0.937 | 1.000 | 0.967 |
| DOSDUR | 186 | 10 | 0 | 9 | 0.949 | 0.954 | 0.951 |
| TRT | 203 | 1 | 0 | 1 | 0.995 | 0.995 | 0.995 |
| SPLRNAM | 199 | 6 | 0 | 0 | 0.971 | 1.000 | 0.985 |
| SPLRLOC | 157 | 40 | 7 | 1 | 0.797 | 0.994 | 0.885 |
| STDIR | 186 | 19 | 0 | 0 | 0.907 | 1.000 | 0.951 |
| EXSTDTC | 44 | 161 | 0 | 0 | 0.215 | 1.000 | 0.353 |
| BW | 204 | 0 | 0 | 1 | 1.000 | 0.995 | 0.998 |
| CL | 204 | 0 | 0 | 1 | 1.000 | 0.995 | 0.998 |
| FW | 177 | 0 | 0 | 28 | 1.000 | 0.863 | 0.927 |
| LB | 205 | 0 | 0 | 0 | 1.000 | 1.000 | 1.000 |
| MA | 157 | 0 | 0 | 48 | 1.000 | 0.766 | 0.867 |
| MI | 158 | 0 | 0 | 47 | 1.000 | 0.771 | 0.871 |
| OM | 199 | 0 | 0 | 6 | 1.000 | 0.971 | 0.985 |
| PP | 115 | 0 | 0 | 90 | 1.000 | 0.561 | 0.719 |
| VS | 167 | 0 | 0 | 38 | 1.000 | 0.815 | 0.898 |
| EG | 170 | 0 | 5 | 30 | 1.000 | 0.850 | 0.919 |
| CV | 136 | 0 | 0 | 69 | 1.000 | 0.663 | 0.798 |
| RE | 204 | 0 | 0 | 1 | 1.000 | 0.995 | 0.998 |

**C** PRINCE baseline

| | TP | FP | TN | FN | Precision | Recall | F1 |
|---|---|---|---|---|---|---|---|
| SPECIES | 39 | 0 | 0 | 0 | 1.000 | 1.000 | 1.000 |
| STRAIN | 38 | 1 | 0 | 0 | 0.974 | 1.000 | 0.987 |
| GLPFL | 34 | 1 | 0 | 4 | 0.971 | 0.895 | 0.932 |
| HOUSEGRP | 33 | 4 | 0 | 2 | 0.892 | 0.943 | 0.917 |
| HUMIDT | 24 | 4 | 0 | 11 | 0.857 | 0.686 | 0.762 |
| TEMP | 27 | 2 | 0 | 10 | 0.931 | 0.730 | 0.818 |
| DIET | 28 | 1 | 0 | 10 | 0.966 | 0.737 | 0.836 |
| ROUTE | 39 | 0 | 0 | 0 | 1.000 | 1.000 | 1.000 |
| EXDOSFRQ | 30 | 1 | 0 | 8 | 0.968 | 0.789 | 0.870 |
| EXDOSTXT | 37 | 1 | 0 | 1 | 0.974 | 0.974 | 0.974 |
| DOSDUR | 26 | 6 | 0 | 7 | 0.812 | 0.788 | 0.800 |
| TRT | 37 | 2 | 0 | 0 | 0.949 | 1.000 | 0.974 |
| SPLRNAM | 36 | 1 | 0 | 2 | 0.973 | 0.947 | 0.960 |
| SPLRLOC | 21 | 14 | 0 | 4 | 0.600 | 0.840 | 0.700 |
| STDIR | 31 | 8 | 0 | 0 | 0.795 | 1.000 | 0.886 |
| EXSTDTC | 9 | 29 | 0 | 1 | 0.237 | 0.900 | 0.375 |
| BW | 35 | 0 | 0 | 4 | 1.000 | 0.897 | 0.946 |
| CL | 35 | 0 | 0 | 4 | 1.000 | 0.897 | 0.946 |
| FW | 28 | 0 | 0 | 11 | 1.000 | 0.718 | 0.836 |
| LB | 36 | 0 | 0 | 3 | 1.000 | 0.923 | 0.960 |
| MA | 34 | 0 | 0 | 5 | 1.000 | 0.872 | 0.932 |
| MI | 34 | 0 | 0 | 5 | 1.000 | 0.872 | 0.932 |
| OM | 34 | 0 | 0 | 5 | 1.000 | 0.872 | 0.932 |
| PP | 29 | 0 | 0 | 10 | 1.000 | 0.744 | 0.853 |
| VS | 29 | 0 | 0 | 10 | 1.000 | 0.744 | 0.853 |
| EG | 35 | 0 | 0 | 4 | 1.000 | 0.897 | 0.946 |
| CV | 26 | 0 | 0 | 13 | 1.000 | 0.667 | 0.800 |
| RE | 38 | 0 | 0 | 1 | 1.000 | 0.974 | 0.987 |

**B** PRINCE + NER

| | TP | FP | TN | FN | Precision | Recall | F1 |
|---|---|---|---|---|---|---|---|
| SPECIES | 205 | 0 | 0 | 0 | 1.000 | 1.000 | 1.000 |
| STRAIN | 204 | 1 | 0 | 0 | 0.995 | 1.000 | 0.998 |
| **AGE** | 160 | 10 | 16 | 19 | 0.941 | 0.894 | 0.917 |
| **AGEU** | 162 | 8 | 16 | 19 | 0.953 | 0.895 | 0.923 |
| GLPFL | 198 | 7 | 0 | 0 | 0.966 | 1.000 | 0.983 |
| HOUSEGRP | 204 | 1 | 0 | 0 | 0.995 | 1.000 | 0.998 |
| **HOUSETYP** | 176 | 3 | 24 | 2 | 0.983 | 0.989 | 0.986 |
| HUMIDT | 181 | 12 | 11 | 1 | 0.938 | 0.995 | 0.965 |
| **HUMIDTU** | 176 | 1 | 23 | 5 | 0.994 | 0.972 | 0.983 |
| **LIGHT** | 180 | 0 | 23 | 2 | 1.000 | 0.989 | 0.994 |
| TEMP | 183 | 11 | 10 | 1 | 0.943 | 0.995 | 0.968 |
| **TEMPU** | 180 | 1 | 20 | 4 | 0.994 | 0.978 | 0.986 |
| DIET | 190 | 3 | 7 | 5 | 0.984 | 0.974 | 0.979 |
| **WATER** | 186 | 3 | 7 | 9 | 0.984 | 0.954 | 0.969 |
| ROUTE | 204 | 0 | 0 | 1 | 1.000 | 0.995 | 0.998 |
| **TRTV** | 202 | 1 | 0 | 2 | 0.995 | 0.990 | 0.993 |
| **EXVAMT** | 166 | 6 | 2 | 31 | 0.965 | 0.843 | 0.900 |
| EXDOSFRQ | 200 | 3 | 0 | 2 | 0.985 | 0.990 | 0.988 |
| EXDOSTXT | 192 | 13 | 0 | 0 | 0.937 | 1.000 | 0.967 |
| **EXDOSFRM** | 177 | 3 | 0 | 25 | 0.983 | 0.876 | 0.927 |
| DOSDUR | 202 | 3 | 0 | 0 | 0.985 | 1.000 | 0.993 |
| TRT | 202 | 2 | 0 | 1 | 0.990 | 0.995 | 0.993 |
| **RECSAC** | 204 | 1 | 0 | 0 | 0.995 | 1.000 | 0.998 |
| SPLRNAM | 202 | 3 | 0 | 0 | 0.985 | 1.000 | 0.993 |
| **SPLRCTRY** | 71 | 5 | 0 | 129 | 0.934 | 0.355 | 0.514 |
| SPLRLOC | 183 | 14 | 7 | 1 | 0.929 | 0.995 | 0.961 |
| **SPLRREG** | 179 | 0 | 16 | 10 | 1.000 | 0.947 | 0.973 |
| **TSTFNAM** | 193 | 6 | 0 | 6 | 0.970 | 0.970 | 0.970 |
| STDIR | 204 | 1 | 0 | 0 | 0.995 | 1.000 | 0.998 |
| EXSTDTC | 188 | 17 | 0 | 0 | 0.917 | 1.000 | 0.957 |
| **EXENDTC** | 185 | 17 | 0 | 3 | 0.916 | 0.984 | 0.949 |
| **MANAESTH** | 39 | 81 | 54 | 31 | 0.325 | 0.557 | 0.411 |
| BW | 205 | 0 | 0 | 0 | 1.000 | 1.000 | 1.000 |
| CL | 205 | 0 | 0 | 0 | 1.000 | 1.000 | 1.000 |
| FW | 203 | 2 | 0 | 0 | 0.990 | 1.000 | 0.995 |
| LB | 205 | 0 | 0 | 0 | 1.000 | 1.000 | 1.000 |
| PP | 199 | 6 | 0 | 0 | 0.971 | 1.000 | 0.985 |
| MA | 201 | 4 | 0 | 0 | 0.980 | 1.000 | 0.990 |
| MI | 202 | 3 | 0 | 0 | 0.985 | 1.000 | 0.993 |
| OM | 204 | 1 | 0 | 0 | 0.995 | 1.000 | 0.998 |
| VS | 200 | 5 | 0 | 0 | 0.976 | 1.000 | 0.988 |
| EG | 205 | 0 | 0 | 0 | 1.000 | 1.000 | 1.000 |
| CV | 197 | 8 | 0 | 0 | 0.961 | 1.000 | 0.980 |
| RE | 205 | 0 | 0 | 0 | 1.000 | 1.000 | 1.000 |

**D** PRINCE + NER

| | TP | FP | TN | FN | Precision | Recall | F1 |
|---|---|---|---|---|---|---|---|
| SPECIES | 39 | 0 | 0 | 0 | 1.000 | 1.000 | 1.000 |
| STRAIN | 37 | 2 | 0 | 0 | 0.949 | 1.000 | 0.974 |
| **AGE** | 23 | 0 | 1 | 15 | 1.000 | 0.605 | 0.754 |
| **AGEU** | 23 | 0 | 1 | 15 | 1.000 | 0.605 | 0.754 |
| GLPFL | 39 | 0 | 0 | 0 | 1.000 | 1.000 | 1.000 |
| HOUSEGRP | 39 | 0 | 0 | 0 | 1.000 | 1.000 | 1.000 |
| **HOUSETYP** | 23 | 4 | 4 | 8 | 0.852 | 0.742 | 0.793 |
| HUMIDT | 33 | 4 | 0 | 2 | 0.892 | 0.943 | 0.917 |
| **HUMIDTU** | 27 | 0 | 4 | 8 | 1.000 | 0.771 | 0.871 |
| **LIGHT** | 28 | 0 | 0 | 11 | 1.000 | 0.718 | 0.836 |
| TEMP | 35 | 2 | 0 | 2 | 0.946 | 0.946 | 0.946 |
| **TEMPU** | 30 | 1 | 1 | 7 | 0.968 | 0.811 | 0.882 |
| DIET | 36 | 1 | 0 | 2 | 0.973 | 0.947 | 0.960 |
| **WATER** | 32 | 0 | 0 | 7 | 1.000 | 0.821 | 0.901 |
| ROUTE | 39 | 0 | 0 | 0 | 1.000 | 1.000 | 1.000 |
| **TRTV** | 37 | 0 | 0 | 2 | 1.000 | 0.949 | 0.974 |
| **EXVAMT** | 33 | 4 | 0 | 2 | 0.892 | 0.943 | 0.917 |
| EXDOSFRQ | 37 | 2 | 0 | 0 | 0.949 | 1.000 | 0.974 |
| EXDOSTXT | 38 | 1 | 0 | 0 | 0.974 | 1.000 | 0.987 |
| **EXDOSFRM** | 33 | 1 | 0 | 5 | 0.971 | 0.868 | 0.917 |
| DOSDUR | 35 | 4 | 0 | 0 | 0.897 | 1.000 | 0.946 |
| TRT | 36 | 3 | 0 | 0 | 0.923 | 1.000 | 0.960 |
| **RECSAC** | 39 | 0 | 0 | 0 | 1.000 | 1.000 | 1.000 |
| SPLRNAM | 38 | 0 | 0 | 1 | 1.000 | 0.974 | 0.987 |
| **SPLRCTRY** | 10 | 2 | 10 | 17 | 0.833 | 0.370 | 0.513 |
| SPLRLOC | 25 | 12 | 0 | 2 | 0.676 | 0.926 | 0.781 |
| **SPLRREG** | 32 | 0 | 5 | 2 | 1.000 | 0.941 | 0.970 |
| **TSTFNAM** | 37 | 1 | 0 | 1 | 0.974 | 0.974 | 0.974 |
| STDIR | 37 | 2 | 0 | 0 | 0.949 | 1.000 | 0.974 |
| EXSTDTC | 32 | 7 | 0 | 0 | 0.821 | 1.000 | 0.901 |
| **EXENDTC** | 35 | 3 | 0 | 0 | 0.921 | 1.000 | 0.959 |
| **MANAESTH** | 4 | 23 | 8 | 4 | 0.148 | 0.500 | 0.229 |
| BW | 38 | 0 | 0 | 0 | 1.000 | 1.000 | 1.000 |
| CL | 38 | 0 | 0 | 0 | 1.000 | 1.000 | 1.000 |
| FW | 38 | 0 | 0 | 0 | 1.000 | 1.000 | 1.000 |
| LB | 38 | 0 | 0 | 0 | 1.000 | 1.000 | 1.000 |
| PP | 37 | 2 | 0 | 0 | 0.949 | 1.000 | 0.974 |
| MA | 38 | 1 | 0 | 0 | 0.974 | 1.000 | 0.987 |
| MI | 38 | 1 | 0 | 0 | 0.974 | 1.000 | 0.987 |
| OM | 39 | 0 | 0 | 0 | 1.000 | 1.000 | 1.000 |
| VS | 37 | 2 | 0 | 0 | 0.949 | 1.000 | 0.974 |
| EG | 38 | 1 | 0 | 0 | 0.974 | 1.000 | 0.987 |
| CV | 34 | 5 | 0 | 0 | 0.872 | 1.000 | 0.932 |
| RE | 39 | 0 | 0 | 0 | 1.000 | 1.000 | 1.000 |

**% of studies per field**

0 — 100

Figure 3 - Extraction performance across Roche metadata fields (N = 202 reports). Stacked bar chart showing absolute counts of true positives (TP), false positives (FP), true negatives (TN), and false negatives (FN) per metadata field (A). Test Facility Name and Test Facility Location include 220 evaluated instances because some studies involved multiple facilities. Grouped bar chart showing accuracy, recall, precision, and F1 score per metadata field, calculated from the counts in A and rounded to the nearest integer (B).

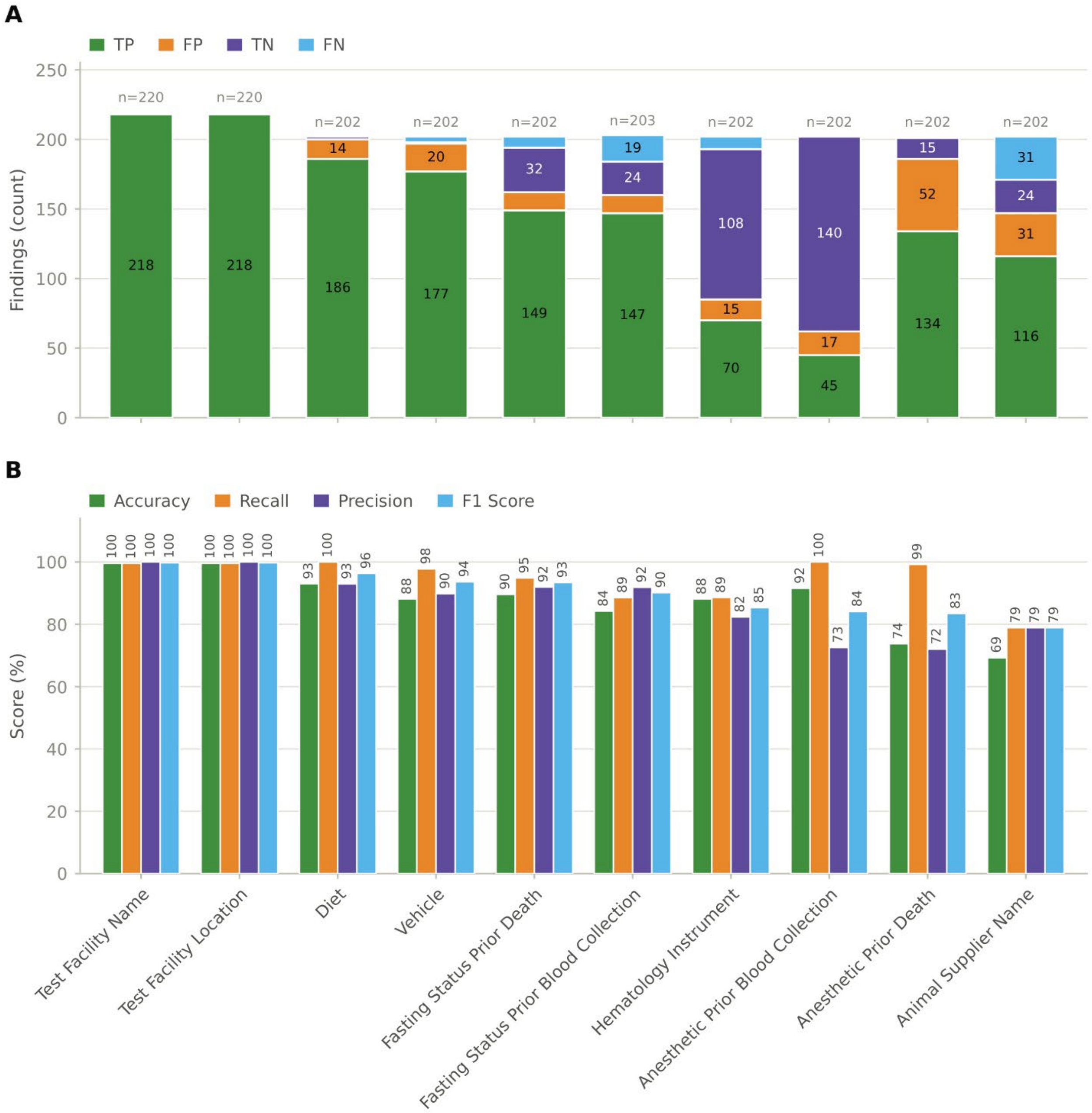

Figure 4 – Fasting status and its impact on clinical pathology parameters. Number of studies and control subjects included in the fasting-status models (A). Counts are shown for analytes with complete panel coverage. Fasting effect on control clinical chemistry in cynomolgus monkey and Wistar Han rat. Per-animal control values (boxes, log scale) and per-study means (black dots) for eight analytes in fasted (green) versus unfasted (orange) studies. Brackets mark the modelled fasting

effect from the study-level mixed model adjusted for sex and specimen matrix (* q<0.05, Benjamini–Hochberg FDR across 16 tests) (B). Modelled fasting effect on control clinical chemistry. Fold-change (fasted/unfasted) with 95 % CI from a per-analyte linear mixed model on log values, with a study random intercept and adjustment for sex and specimen matrix; one panel per species, analytes sorted by effect size. Labels give the percentage change. Red = significant after Benjamini–Hochberg FDR correction across the 16 species × analyte tests (C).

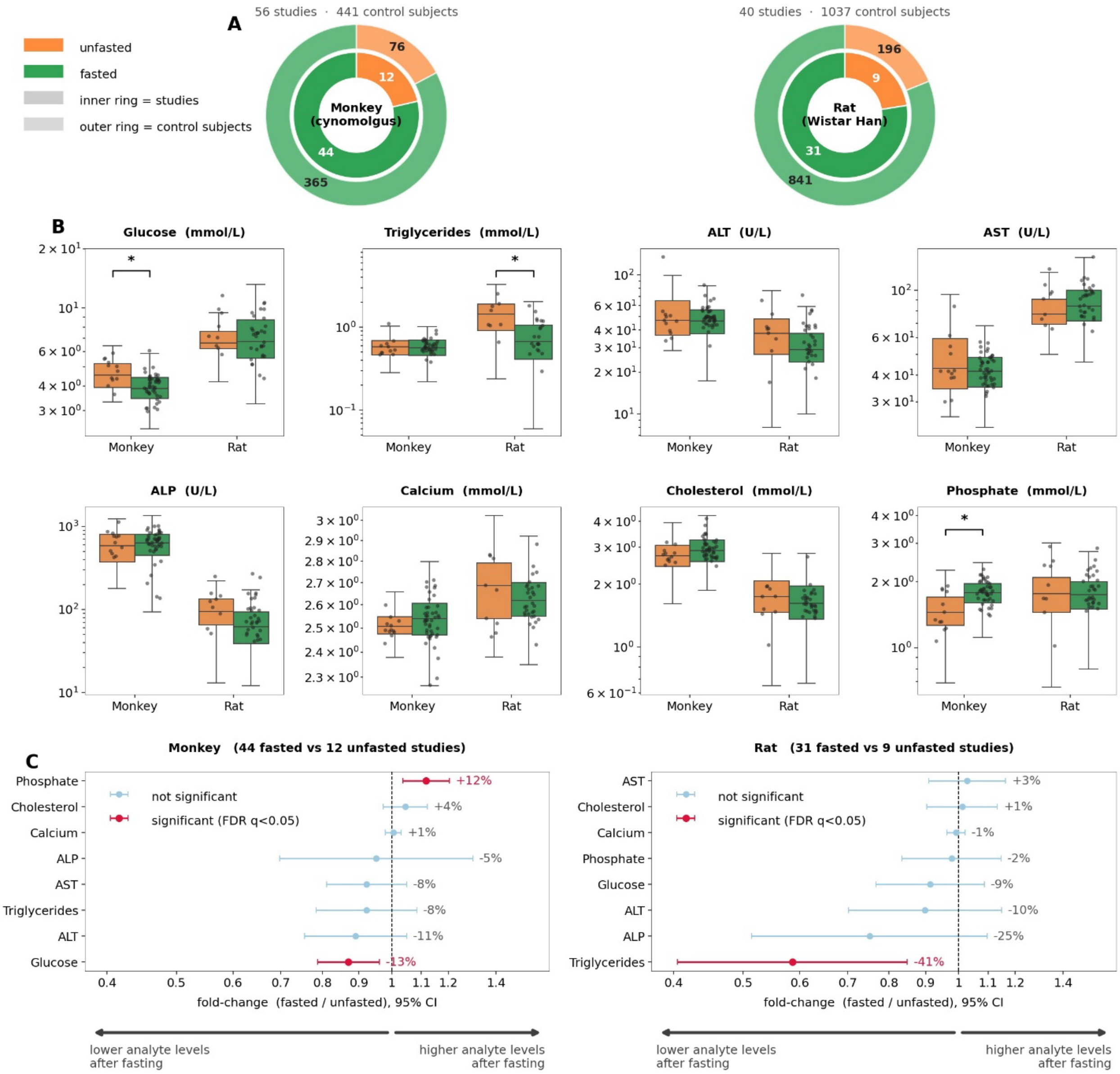

Table 1 - Examples of confusion-matrix categories from study reports. “No clear answer” denotes a standardized missing-value response indicating that no value was extracted.

| **Category** | **Extracted value** | **Manually curated reference value** |
|---|---|---|
| True Positive (full) | Anesthetic = carbon dioxide inhalation | Anesthetic = carbon dioxide inhalation |
| True Positive (partial) | Route of administration = Oral | Route of administration = Oral Gavage |
| False Positive (no ground truth) | Fasting status = fasted | Fasting status = Not Available |
| False Positive (wrong value) | Animal diet = Kliba Nafag #3436 | Animal diet = Provimi Kliba 3437 |
| True Negative | Animal supplier = "No clear answer" | No value available in the report |
| False Negative | Animal supplier = "No clear answer" | Animal supplier = Ellegaard Göttingen Minipigs A/S, DK-4261 Dalmose, Denmark |

**Supplementary Materials**

**Supplementary Material 1**

""""

**Evaluation Criteria:**

1. **Correctness**: An answer is "correct" if it matches the reference answer. Variations in wording, order, expression as a sentence or as a single expression, naming conventions, or minor typos are acceptable as long as the meaning remains accurate in the given context. Numbers can be given as integers or floating point values (e.g. 2 or 2.0) and both are correct. Further specifications are added in the Details later on. Empty answers, "Not Found", "Not Reported", and "Not Available" should all be considered equivalent answers, but not "Not applicable".

2. **Partial Correctness**: An answer is "partially correct" if it accurately addresses the question, even if expressed differently, but does not exactly match the reference answer. Here are some examples:

- An address missing the city but containing the correct country is a partially correct answer;

- A date containing the correct year and month but not the correct day, should be considered incorrect. Therefore, dates cannot be considered partially correct and must always be correct or incorrect;

- If time is expressed in number of days, weeks, months or years, approximates are accepted. For example, '28 days', '32 days' or '4 weeks' are all similar answers that should be considered partially correct. Or another example, '91 days', '95 days' or '13 weeks' should also be considered partially correct answers;

- "ORAL" may be used to express a route of administration via diet (dietary), which in this case indicates a correct answer, but if "ORAL" was used to indicate an "ORAL GAVAGE" route of administration than the evaluation is that the answer was partially correct. Similarly, "intravenous" is only partially correct answer for reference answers "intravenous bolus" and "intravenous infusion";

- Regarding route of administration, intragastric means the same as oral gavage;

- "One time" or "single dose" maybe be considered synonyms to "ONCE" when evaluating dosage frequency;

- "Metal boxes" are synonym for "conventional cages" when evaluating animal housing type;

- "Kolliphor" and "Solutol" are synonyms components of administration vehicles. Also, the order in which the components of a chemical solution are described do not change the composition of the solution;

- When comparing codes like "BAY 123456", the presence of a dash does not change the code, meaning "BAY 123-456" would be an equivalent;

- "Single housed" or "singly housed" are synonyms, and other variations of the terminology are acceptable:

- "Filtered water" and "Tap water" are synonyms;

- If a number range is given, consider them partially correct if the answer ranges overlap. For example, "19-24" overlaps with "22-25" but not with "25-27";

- A person's name may be spelled in full or reduced. For example, "Carlos Vieira-Vieira", "Dr. Carlos Vieira-Vieira", "Dr. C. Vieira-Vieira" are all entries identifying the same person and therefore should be considered as the same;

- Numbers may be expressed as integers or floats and should be considered as the same number. For example, "2017" and "2017.0" should be considered as the same number, and therefore correct;

3. **Imprecision:** Any answer that does not accurately address the question, contains significant misunderstandings or is not reflecting all important aspects of the reference answer, should be marked as "Incorrect".

4. **Additional Considerations:**

Technical Terminology: Ensure your account for relevant technical terms in toxicology and biology.

Order of Information: The order of items in answers does not affect precision, especially for categories where multiple entries may exist. Multiple answers are often divided by semicolons or structured as an array.

**Details:**

- **Category and Explanation**: {field} : {flat_field_descriptions[field]}

- **Student's Answer**: {extracted_answer}

- **Reference Correct Answer**: {curated_answer}

**Instructions:**

- Compare the student's answer to the correct answer based on the criteria above.

- Your reply should include only your final assessment: correct, partially correct or incorrect.

"""

**Supplementary Material 2**

"""

The strain of the animals used in this study. Typical strains are the following: GOTTINGEN, CYNOMOLGUS, WISTAR WU, SPRAGUE-DAWLEY, RNU, NMRI, DOMESTIC SHORT HAIR, B6C3F1, CD1(ICR), CB6F1-TGN (RASH2), RasH2, SPRAGUE DAWLEY (SD), CRL:CD, WISTAR, BEAGLE, MARMOSET, CHIMPANZEE, NEW ZEALAND, WISTAR HAN, BABOON, RHESUS, ZUCKER, LONG EVANS. Only include the information on the strain, as numbers or other information are not required. Specifically, check in the description of the study for the animal in question and which strain of it was used. The information on the strain may not be always found in the text. In that case, one can assume the strain based on the species of animals used, but first please make sure that the strain information is not present in the text. If pig or minipigs were used, assume GOTTINGEN. If dog was used, assume BEAGLE. If monkey or non-human primate was used, assume CYNOMOLGUS. If rabbit was used, assume NEW ZEALAND. If mouse was used, assume CD1(ICR). If rat was used, assume WISTAR.

## Further instructions:

The value you extract from the text will likely need to be harmonized before providing the final answer. Below is a json file containing a mapping relationship that you should use. This json contains many examples of harmonized answers (keys in the dictionary) and the values that have been extracted from the text (values in the dictionary). Use it to learn and provide the correctly harmonized final answer.

{

"BEAGLE": ["Beagle", "BEAGLE"],

"CD1(ICR)": ["Crl:CD-1(ICR)BR", "Crl:CD-1 (ICR)BR", "Crl:CD1 (ICR)", "Crl:CD1(Icr)", "Crl:CD1(ICR)BR", "Crl:CD1 (Icr)", "CD1(ICR)"],

"CYNOMOLGUS": ["CYNOMOLGUS"],

"NEW ZEALAND": ["NEW ZEALAND"],

"SPRAGUE-DAWLEY": ["Sprague Dawley SD"],

"WISTAR": ["Wistar", "WISTAR"],

"WISTAR HAN": ["RccHanWist", "Crl WI:Han", "Crl:Wi(Han)", "RccHan:WIST", "HsdRCCHan:Wist", "Crl:WI(Han)", "Crl:WI (Han)", "RccHanWIST", "WISTAR HAN", "Crl: WI (Han)"],

"WISTAR WU": ["HsdCpb:WU [SPF]"],

"CB6F1-TGN (RASH2)": ["RasH2 Mice", "RasH2"]

}

"""

**Supplementary Material 3**

Report-core identification used the table of contents (TOC) where available. When automatic TOC detection failed, the initial report pages were extracted and parsed to identify TOC content, and the trigger terms “reference,” “archiving,” and “signature” were used to identify the final page of the report core. Parsing was restricted to pages up to this endpoint. If no reliable endpoint could be identified, the first 100 pages were retained. This occasionally included appendix material but was preferred over potentially excluding information required for extraction.

Parser selection depended on computational resources available at runtime. MinerU required GPU execution, whereas olmOCR supported both CPU and GPU execution.

The extraction schema followed a class-based Python design in which related metadata fields were grouped by topic (Supplementary Figure 1). Each field specified its extraction instructions and required output format. Study-specific extension fields captured information not consistently represented in available structured metadata but considered relevant to cross-study comparison and VCG development. Anesthesia before blood collection was included, for example, because anesthesia protocols can substantially affect clinical-pathology measurements and therefore historical-control comparability (Gurjanov et al., 2023).

All metadata fields, corresponding SEND variables or domains where applicable, and extraction definitions are provided in Supplementary Table 1. If a field could not be extracted, the pipeline returned a standardized missing-value response. For GLP status, recovery period, and SEND-domain measurement availability, absence of evidence was instead encoded as a negative value.

**Supplementary Table 1**

Metadata fields selected for extraction from preclinical toxicity study reports.

| **Category** | **Metadata field** | **SEND variable/domain** | **Extraction definition used in this study** | **Missing-value instruction** |
|---|---|---|---|---|
| Animal characteristics | Animal species | SPECIES | Species of the test animals | Not reported |
| Animal characteristics | Strain | STRAIN | Strain of the test animals | Not reported |
| Animal characteristics | Age at start of experiment | AGE, AGEU | Age of animals at the start of the experimental phase, including unit | Not reported |
| Husbandry | Basal diet | DIET | Diet given to study animals | Not reported |
| Husbandry | Drinking water | WATER | Water provided to study animals | Not reported |
| Study conduct | Good laboratory practice flag | GLPFL | GLP conditions | Negative if absent |
| Housing | Housing group | HOUSEGRP | Animal grouping or housing group | Not reported |
| Housing | Cage type | HOUSETYP | Cage used for animal housing | Not reported |
| Housing | Humidity | HUMIDT, HUMIDTU | Reported humidity conditions, including unit | Not reported |
| Housing | Light cycle | LIGHT | Reported light/dark cycle | Not reported |
| Housing | Temperature | TEMP, TEMPU | Reported room or housing temperature, including unit | Not reported |
| Supplier | Animal supplier country | SPLRCTRY | Country of the animal supplier | Not reported |
| Supplier | Animal supplier location | SPLRLOC | Location of the animal supplier | Not reported |
| Supplier | Animal supplier name | SPLRNAM | Name of the entity that supplied or sourced the animals | Not reported |
| Supplier | Animal supplier region | SPLRREG | Region of the animal supplier | Not reported |
| Test facility | Test facility name | TSTFNAM | Name of the facility where the animal study was conducted | Not reported |
| Test facility | Test facility location | Not applicable | City and country where the test facility is located | Not reported |
| Study personnel | Study director | STDIR | Name of the study director | Not reported |
| Treatment and dosing | Treatment | TRT | Test article administered to the animals | Not reported |
| Treatment and dosing | Treatment vehicle | TRTV | Substance used as carrier or solvent for the test compound | Not reported |
| Treatment and dosing | Route of administration | ROUTE | Route by which the test article or control was administered | Not reported |
| Treatment and dosing | Volume of administration | EXVAMT | Administered dose volume | Not reported |
| Treatment and dosing | Dosing frequency | EXDOSFRQ | Frequency of test article or control administration | Not reported |
| Treatment and dosing | Treatment dosing groups | EXDOSTXT | Description of treatment or dose groups | Not reported |
| Treatment and dosing | Treatment formulation | EXDOSFRM | Formulation of the administered treatment | Not reported |
| Treatment and dosing | Dosing duration | DOSDUR | Duration of treatment or dosing period | Not reported |
| Study design | Recovery period | RECSAC | Whether animals were assigned to a recovery period or recovery sacrifice group | Negative if absent |
| Exposure timing | Exposure start date | EXSTDTC | Start date of treatment or exposure | Not reported |
| Exposure timing | Exposure end date | EXENDTC | End date of treatment or exposure | Not reported |
| Measurement domain availability | Body weight measurements | BW | Whether body weight measurements were reported | Negative if absent |
| Measurement domain availability | Clinical observations | CL | Whether clinical observations were reported | Negative if absent |
| Measurement domain availability | Food and water intake | FW | Whether food and/or water intake measurements were reported | Negative if absent |
| Measurement domain availability | Laboratory test results | LB | Whether laboratory test results were reported | Negative if absent |
| Measurement domain availability | Pharmacokinetic parameters | PP | Whether pharmacokinetic parameters were reported | Negative if absent |
| Measurement domain availability | Macroscopic findings | MA | Whether macroscopic pathology findings were reported | Negative if absent |
| Measurement domain availability | Microscopic findings / histopathology | MI | Whether microscopic pathology or histopathology findings were reported | Negative if absent |
| Measurement domain availability | Organ measurements | OM | Whether organ weight or organ measurement data were reported | Negative if absent |
| Measurement domain availability | Vital signs | VS | Whether vital signs were reported | Negative if absent |
| Measurement domain availability | ECG test results | EG | Whether electrocardiography test results were reported | Negative if absent |
| Measurement domain availability | Cardiovascular test results | CV | Whether cardiovascular test results were reported | Negative if absent |
| Measurement domain availability | Respiratory test results | RE | Whether respiratory test results were reported | Negative if absent |
| Anesthesia | Anesthetic before blood collection | Study-specific; proposed label: MANAESTH_BLOOD | Anesthetic administered before blood collection | Not reported |
| Anesthesia | Anesthetic before euthanasia | Study-specific; proposed label: MANAESTH_DEATH | Anesthetic administered before euthanasia or sacrifice | Not reported |
| Fasting | Fasting status before blood collection | Study-specific | Fasting status of animals before blood sampling for serum chemistry or hematology | Not reported |
| Fasting | Fasting status before sacrifice or necropsy | Study-specific | Fasting status of animals before sacrifice, euthanasia, or necropsy preparation | Not reported |
| Hematology | Hematology instrument | Study-specific | Name or model of the analyzer used for hematology analyses | Not reported |

**Supplementary Figure 1**

Example of the prompt formulation for the Test Facility extraction item and its structure as an object.

PYDANTIC SCHEMA

**TestFacilityItem**

Extracted value | Source page | Verbatim evidence

| FIELD NAME RETURNED | TYPE | DESCRIPTION GIVEN TO THE MODEL |
|---|---|---|
| TestFacilityName | str | The name of the facility where the test with animals are conducted. |
| TestFacilityNamePage | str | The page from which you extracted the TestFacilityName as '<Page Number>'. |
| TestFacilityLocation | str | A city and a country where the test facility is located. |
| TestFacilityLocationPage | str | The page from which you extracted the TestFacilityLocation as '<Page Number>'. |
| ReportText | str | Quote the actual report text (<report_text>) based on which you extracted the information for this finding. |
| Header | str | Provide the header of the report section in which the ReportText (<report_text>) is contained. |

## Supplementary Figure 2

Per-field distribution of extraction outcomes benchmarked against manually curated metadata fields. Stacked bar plots are shown for: Training set (n = 205), PRINCE baseline (A); Training set (n = 205), PRINCE plus extraction pipeline combined output after rule-based merging (B); Validation set (n = 39), PRINCE baseline (C); and validation set (n = 39), PRINCE plus extraction pipeline combined output after rule-based merging (D). In combined panels (B and D), bold field labels indicate metadata fields not present in PRINCE.

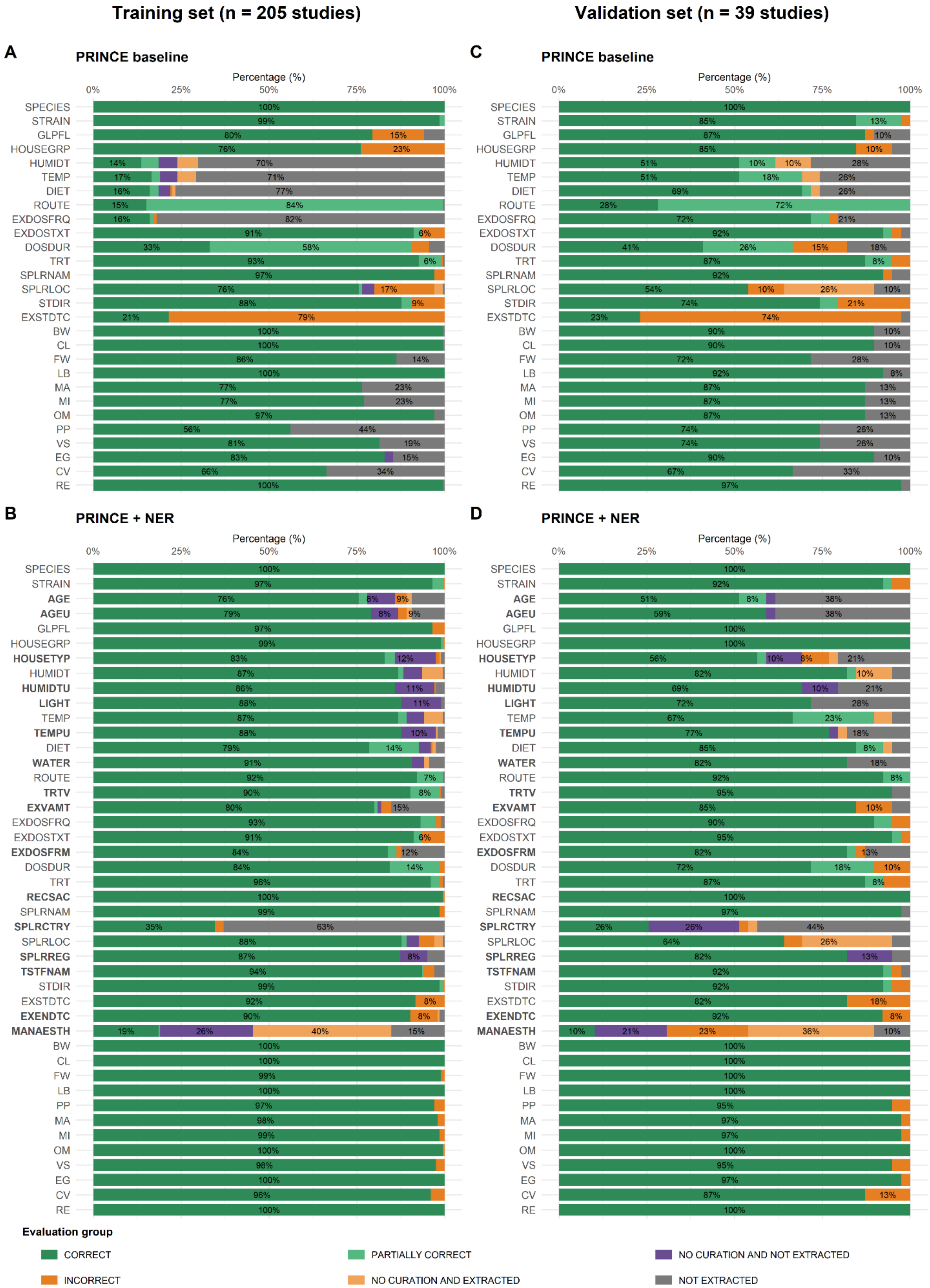

Training set (n = 205 studies)
Validation set (n = 39 studies)
A
PRINCE baseline
C
PRINCE baseline
Percentage (%)
0%
25%
50%
75%
100%
SPECIES
STRAIN
GLPFL
HOUSEGRP
HUMIDT
TEMP
DIET
ROUTE
EXDOSFRQ
EXDOSTXT
DOSDUR
TRT
SPLRNAM
SPLRLOC
STDIR
EXSTDTC
BW
CL
FW
LB
MA
MI
OM
PP
VS
EG
CV
RE
B
PRINCE + NER
D
PRINCE + NER
SPECIES
STRAIN
AGE
AGEU
GLPFL
HOUSEGRP
HOUSETYP
HUMIDT
HUMIDTU
LIGHT
TEMP
TEMPU
DIET
WATER
ROUTE
TRTV
EXVAMT
EXDOSFRQ
EXDOSTXT
EXDOSFRM
DOSDUR
TRT
RECSAC
SPLRNAM
SPLRCTRY
SPLRLOC
SPLRREG
TSTFNAM
STDIR
EXSTDTC
EXENDTC
MANAESTH
BW
CL
FW
LB
PP
MA
MI
OM
VS
EG
CV
RE
Evaluation group
CORRECT
PARTIALLY CORRECT
NO CURATION AND NOT EXTRACTED
INCORRECT
NO CURATION AND EXTRACTED
NOT EXTRACTED